\documentclass[fleqn,twoside]{article}
\usepackage[headings]{espcrc2}
\usepackage{epsf}
\usepackage{amssymb,latexsym}

\usepackage[dvips]{graphicx}
\usepackage[figuresright]{rotating}
\newcommand{\mev}{\mbox{\rm MeV}}
\newcommand{\AmS}{{\protect\the\textfont2
  A\kern-.1667em\lower.5ex\hbox{M}\kern-.125emS}}
\newcommand{\lsim}{\stackrel{<}{_\sim}}

\title{Analysis of $\tau^- \rightarrow K_S \pi^- \nu_{\tau}$ Belle data in a chiral
       framework \thanks{Report IFIC/08-51. Talk given at the 14th International 
       QCD Conference, Montpellier (France), 7-12th July 2008.
       This work has been supported in part by the EU MRTN-CT-2006-035482 (FLAVIAnet),
by MEC (Spain) under grant FPA2007-60323 and by the Spanish Consolider-Ingenio 2010 Programme
CPAN (CSD2007-00042).}}

\author{J.~Portol\'es\address[MCSD]{IFIC, Universitat de Val\`encia - CSIC, \\
                                    Apt. Correus 22085, E-46071 Val\`encia, Spain}}
       
\runtitle{}
\runauthor{}

\begin{document}

\begin{abstract}
The recent measurement of the $\tau^- \rightarrow K_S \pi^- \nu_{\tau}$ spectrum by the Belle
Collaboration provides the frame to set forth a theoretical description of the decay, which is
based on the contributing vector $F_{+}^{K \pi}(s)$ and scalar $F_0^{K \pi}(s)$ form factors.
We show that a good representation of data is obtained through the use of form factors 
calculated within resonance chiral theory and constrained by dispersion relations and short-distance
QCD. Hence we obtain a determination of $K^*(892)$ parameters and the low-energy parameterization 
of $F_{+}^{K \pi}(s)$.
\vspace{1pc}
\end{abstract}

\maketitle

\section{INTRODUCTION}

Hadronic decays of the $\tau$ lepton provide an excellent setting for the study of 
hadronization of vector and axial-vector QCD currents \cite{bnp92,np88,pich89}
at $E \lsim M_{\tau}$. Moreover the experimental separation of the Cabibbo-allowed 
and Cabibbo-suppressed modes into strange particles \cite{dhz05,opal04,aleph99} has also 
opened a way out for the determination of $|V_{us}|$ \cite{gjpps07,gjpps04} or the 
mass of the strange quark \cite{bck04}, from the $\tau$ strange
spectral function.
\par
Exclusive hadronic $\tau$ decays bring in properties related with the hadronization procedure, 
that provide valuable information on the description of QCD in the low-energy region dominated by 
light-flavoured resonances. The dominant contribution to the Cabibbo-suppressed $\tau$ decay
widths arises from the decay $\tau \rightarrow K \pi \nu_{\tau}$. Though its distribution
function has been measured in the past \cite{opal04,aleph99}, high-statistics data for the 
spectrum from Belle \cite{belle07} and for the total branching fraction from BaBar
\cite{babar07} are now available. 
\par
Though earlier attempts to describe this channel have been tried \cite{fm96} within an 
ad-hoc modelization with Breit-Wigner functions, 
in Ref.~\cite{jpp06} we have tackled the description of the $\tau \rightarrow K \pi \nu_{\tau}$
data through a thorough construction of the relevant form factors by imposing the model-independent constraints of dispersion relations, chiral symmetry and their asymptotic QCD behaviour. Later on 
\cite{Jamin:2008qg} we have applied that study to the analysis of Belle data \cite{belle07}. Here
I collect the summary of our results. Other recent works have also approached the study of this channel
\cite{mou07}.
\par
The relevant form factors are defined by the hadronic matrix element~:
\begin{eqnarray}
\langle \, \pi^-(p') \, \overline{K}^0(p) | \, \overline{s} \, \gamma_{\mu} \, u \, | 0 \rangle =  && 
\nonumber \\
\left[ \frac{Q_{\mu} Q_{\nu}}{Q^2} - g_{\mu \nu} \right] && \! \! \! \! \! \! \! \! \! \! \! \!\! \!
\!\! \! \!\! \!\left( p - p' \right)^{\nu} F_{+}^{K \pi}(Q^2) \nonumber \\
 -  \frac{\Delta_{K \pi}}{Q^2} \, Q_{\mu} \,  F_0^{K \pi}& \! \! \! \!\! \! (Q^2) \, , &  
\end{eqnarray}
where $Q_{\mu} = (p+p')_{\mu}$ and $\Delta_{K \pi} = M_K^2-M_{\pi}^2$. Here $F_+^{K \pi}(Q^2)$ and 
$F_0^{K \pi}(Q^2)$ are the vector and scalar $K \pi$ form factors, respectively. The general expression
for the $\tau^- \rightarrow K_S \pi^- \nu_{\tau}$ differential decay distribution is given by~:
\begin{eqnarray}
\label{eq:diffiero}
\frac{d \Gamma_{K_S \pi}}{d \sqrt{s}} = \frac{G_F^2 |V_{us}|^2 M_{\tau}^3}{96 \, \pi^3 \, s} \, S_{\mbox{\scriptsize{EW}}} 
 \left( 1- \frac{s}{M_{\tau}^2} \right)^2   \nonumber  \\ \times
\Bigg[ \left( 1+2\frac{s}{M_{\tau}^2} \right) q_{K \pi}^3 \left| F_{+}^{K \pi}(s) \right|^2   & & \nonumber \\  \! \! \! \! \!\! \! \! \! \!
+ \, \frac{3 \Delta_{K \pi}^2}{4 \, s} \, q_{K \pi}  \left| F_0^{K \pi}(s) \right|^2 \Bigg] , &&
\end{eqnarray}
being $q_{K \pi}$ the kaon momentum in the rest frame of the hadronic system, $s=Q^2$ and $S_{\mbox{\scriptsize{EW}}}$
is an electro-weak correction factor.

\section{THE FORM FACTORS}

In Ref.~\cite{jpp06} we have studied a theoretical representation of the vector form factor $F_+^{K \pi}(s)$ in complete analogy to the description of the pion form factor presented in Refs.~\cite{gp97}.
This approach includes our present knowledge on phenomenological hadronic Lagrangians, short-distance QCD, the large-$N_C$ expansion as well as analyticity and unitarity. 
\par
The dynamical information of the vector form factor is dominantly carried out by the lightest s-flavoured vector resonance, namely $K^* = K^*(892)$. Since the $\tau$ lepton can also decay hadronically into the second vector resonance $K^{*'} = K^*(1410)$, we have also included it in our parameterization. Its expression
is given by~:
\begin{eqnarray}
\label{eq:our}
F_+^{K\pi}(s) \,=\, \Bigg[\, \frac{M_{K^*}^2+\gamma\,s}{M_{K^*}^2 - s -
iM_{K^*}\Gamma_{K^*}(s)} \Bigg.  && \nonumber \\
 \Bigg. - \frac{\gamma\,s}{M_{K^{*'}}^2 - s -
iM_{K^{*'}}\Gamma_{K^{*'}}(s)} \,\Bigg]  \times \!\! \! \! \! \!  \!\! \! \! \! \! \! \! \!&& \nonumber \\ 
\!\! \! \! \! \!  \!\! \! \! \! \! \! \!\exp \left\{ \frac{3}{2}
\mbox{Re} \left[ \widetilde{H}_{K\pi}(s)+
\widetilde{H}_{K\eta}(s) \right] \right\}  &\!\! \! \! .& 
\end{eqnarray}
This incorporates all known constraints from Chiral Perturbation Theory \cite{gl85} and Resonance Chiral Theory (R$\chi$T) \cite{egpr89,eglpr89}. The relation of the parameter $\gamma$ to the 
R$\chi$T couplings is given by $\gamma = F_V G_V / (F_K F_{\pi}) -1$ when we impose the vanishing of the
$N_C \rightarrow \infty$ form factor at $s \rightarrow \infty$. The expressions of the vector resonances  off-shell widths can be seen in Ref.~\cite{Jamin:2008qg}. In Eq.~(\ref{eq:our}) $\widetilde{H}(s) \equiv
H(s) - 2 L_9^r s / (3 F_K F_{\pi})$ and $H(s)$ can be read from Ref.~\cite{gl85b}. In the following we
will call our proposal in Eq.~(\ref{eq:our}) as the {\em Chiral form for} $F_+^{K \pi}(s)$.
\par
For the sake of comparison we have also considered de vector form factor constructed as combinations of 
Breit-Wigner functions \cite{fm96}~:
\begin{eqnarray} \label{eq:theirs}
 \frac{F_{+}^{K \pi}(s)}{F_+^{K \pi}(0)} & = & \frac{BW_{K^*}(s) + \beta BW_{K^{*'}}(s)}{1+\beta} \, ,
\nonumber \\
BW_{R}(s) & \equiv & \frac{M_{R}^2}{M_{R}^2-s-i M_{R} \Gamma_{R}(s)} \, ,
\end{eqnarray}
where in the width of the $K^*$ only the $K \pi$ contribution is included. Hence $\beta=0$ includes only
the $K^*(892)$ intermediate state.
\par
The scalar form factor $F_0^{K \pi}(s)$ was calculated in Refs.~\cite{jop00} in the
framework of R$\chi$T with the implementation of constraints from dispersion theory as well as the short-distance 
QCD ruled behaviour of the form factor. We will use thoroughly the results of these studies. To proceed 
we will take the central value of the scalar form factor and will not consider its error. See Ref.~\cite{Jamin:2008qg} for a discussion on this issue.
\par
A still hotly debated issue is the value of the form factors at $s=0$ where $F_+^{K \pi}(0) = F_0^{K \pi}(0)$
\cite{Cirigliano:2005xn}. However from Eq.~(\ref{eq:diffiero}) we notice that the normalization is 
given by $|V_{us}| F_+^{K \pi}(0)$. Hence we only need to determine the shape of normalized (at the
origin) form factors. From the analyses of semi-leptonic kaon decays we will take \cite{Antonelli:2008jg}~:
\begin{equation}
 |V_{us}| F_+^{K^0 \pi^-}(0) \, = \, 0.21664 \pm 0.00048 \, .
\end{equation}
Then we will be able to predict the total branching fraction $B\left[ \tau^- \rightarrow K_S \pi^- \nu_{\tau} \right]$ from a fit of the shape of the form factors, independent of the normalisation problem.

\section{FITS TO THE BELLE SPECTRUM}

In order to analyse the Belle data \cite{belle07} we take the following Ansatz~:
\begin{equation}
 0.0115 [\mbox{GeV}/\mbox{bin}] \, \frac{{\cal N}_T}{\Gamma_{\tau} \overline{B}_{K \pi}} \, 
\frac{\mbox{d} \Gamma_{K_S \pi}}{\mbox{d} \sqrt{s}} \, ,
\end{equation}
The $11.5 \, \mbox{MeV}$ was the bin-width chosen by the experimentalists, ${\cal N}_T = 53110$ is the total
number of observed signal events, $\Gamma_{\tau}$ is the total decay width of the $\tau$ lepton and 
$\overline{B}_{K \pi}$ is a normalisation factor. A perfect agreement between data and the fit function
would imply that $\overline{B}_{K \pi} = B[\tau^- \rightarrow K_S \pi^- \nu_{\tau}]$. Then we proceed to
analyse different fits. A discussion of 
the numerical input for the vector form factors is accounted for in Ref.~\cite{Jamin:2008qg}.
\par
\begin{figure}[!th]
\hspace*{-0.1cm}
\includegraphics*[scale=0.31,clip]{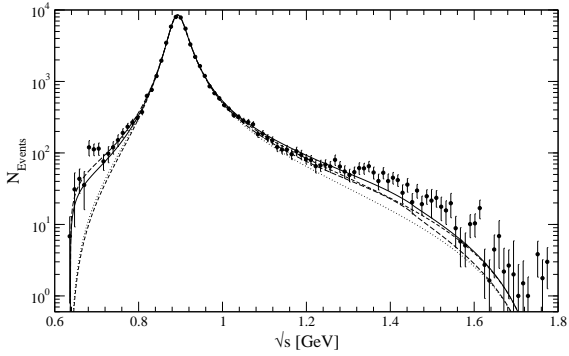}
\vspace*{-0.8cm}
\caption{\label{fig:1} Fit result for the differential decay distribution of the decay $\tau^- \rightarrow K_S \pi^- \nu_{\tau}$, when fitted with a pure $K^*$ vector resonance [dotted (Breit-Wigner) and short-dashed (chiral) curves] or with
$K^*$ plus the scalar form factor (long-dashed and solid curves, respectively). }
\vspace*{-0.5cm}
\end{figure}

In our first fit we employ the Belle data in the range $0.808 - 1.015 \, \mbox{GeV}$ where the vector form
factor dominates. The result is shown in the Fig.~\ref{fig:1} for both our chiral form factor 
(\ref{eq:our}) with $\gamma=0$ and the Breit-Wigner expression (\ref{eq:theirs}) with $\beta=0$. 
In both cases we find $\chi^2/ \mbox{n.d.f.} \simeq 1.9$.
\par
Next we include the scalar form factor $F_0^{K \pi}$ and extend the fitted energy region within\footnote{The data points 5,6 and 7 in the range $0.682 -0.705  \, \mbox{GeV}$ where excluded though.}
$0.636 - 1.015 \, \mbox{GeV}$. Now the goodness of the fit for both vector form factors stays around
$\chi^2 / \mbox{n.d.f.} \simeq 1.6$ and the curves are, again, given in Fig.~\ref{fig:1}. As can be seen
the inclusion of the scalar form factor is all-important in the low-energy region up to $\sqrt{s} \sim
0.9 \, \mbox{GeV}$ where it dominates over the vector one.
\par
\begin{figure}[!th]
\hspace*{-0.1cm}
\includegraphics*[scale=0.31,clip]{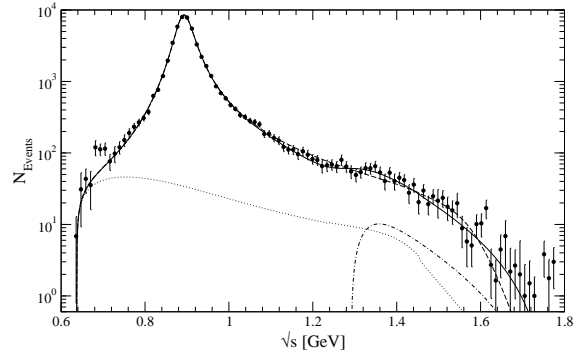}
\vspace*{-0.8cm}
\caption{\label{fig:2} Main fit result to the Belle data \cite{belle07} for the differential decay distribution of the
decay $\tau^- \rightarrow K_S \pi^- \nu_{\tau}$. Our theoretical description includes the Breit-Wigner 
(dashed line) or Chiral (solid line) vector form factors with two resonances, as well as the scalar form
factor. The scalar (dotted line) and $K^{*'}$, in the Chiral case, (dashed-dotted) contributions are also shown.}
\vspace*{-0.5cm}
\end{figure}

Finally we fit the full energy range (but for the three points quoted in footnote 2) by including the 
$K^{*'}$ in the vector form factors. The results of the fit are collected in Tab.~\ref{tab:1}, where we also show the results for the normalization $\overline{B}_{K \pi}$. The corresponding spectra are displayed in 
Fig.~\ref{fig:2}. It is relevant to notice that, over the full range, the better agreement with Belle data is achieved by our chiral vector form factor (\ref{eq:our}) though for both we have $\chi^2 / \mbox{n.d.f.} 
\simeq 1$.  After a thorough study of the error sources \cite{Jamin:2008qg} we obtain~:
\begin{equation}
 B \left[ \tau^- \rightarrow K_S \pi^- \nu_{\tau} \right] \, = \,
(0.427 \pm 0.024) \% \, ,
\end{equation}
in good agreement with the value obtained by the Belle Collaboration \cite{belle07}
$B \left[ \tau^- \rightarrow K_S \pi^- \nu_{\tau} \right]  = (0.404 \pm 0.013) \%$.

\begin{table*}[!ht]
\renewcommand{\arraystretch}{1.1}
\begin{center}
\begin{tabular}{crr}
\hline
 & BW form for $F_+^{K\pi}(s)$ & Chiral form for $F_+^{K\pi}(s)$ \\
\hline
$B_{K\pi}$ $(\%)$ & $0.421$ & $0.427$ \\ 
$\bar B_{K\pi}$ $(\%)$ &
$0.423 \pm 0.012$ & $0.430 \pm 0.011$ \\
$M_{K^*} \, (\mev)$ & $895.12 \pm 0.19$ & $895.28 \pm 0.20$ \\
$\Gamma_{K^*} \, (\mev)$ & $46.79 \pm 0.41$ & $47.50 \pm 0.41$ \\
$M_{K^{*'}}\, (\mev)$ & $1598 \pm 25$ & $1307 \pm 17$ \\
$\Gamma_{K^{*'}} \, (\mev)$ & $224 \pm 47$ & $206 \pm 49$ \\
$\beta$, $\gamma$ & $-\,0.079 \pm 0.010$ & $-\,0.043 \pm 0.010$ \\
\hline
$\chi^2/$n.d.f. & 88.7/81 & 79.5/81 \\
\hline
\end{tabular}
\end{center}
\caption{Full fit to the Belle $\tau^- \rightarrow K_S \pi^-\nu_\tau$ spectrum with the two
$K^*$ and $K^{*'}$ vector resonances in $F_+^{K\pi}(s)$ and the central value of the scalar
form factor $F_0^{K\pi}(s)$.\label{tab:1}}
\vspace*{-0.3cm}
\end{table*}

\section{SLOPE AND CURVATURE OF $F_+^{K \pi}(s)$}

The general expansion of the vector form factor around $s=0$ is given by~:
\begin{equation}
\frac{F_+^{K \pi}(s)}{F_+^{K \pi}(0)} =  1 + \lambda_+' \frac{s}{M_{\pi^-}^2} + 
\frac{\lambda_+''}{2} \frac{s^2}{M_{\pi^-}^4} + \frac{\lambda_+'''}{6}  
\frac{s^3}{M_{\pi^-}^6} + ...
\end{equation}
where $\lambda_+'$, $\lambda_+''$ and $\lambda_+'''$ are the slope, curvature and cubic expansion
parameters, respectively. From the results in Tab.~\ref{tab:1} we find~:
\begin{eqnarray}
 \lambda_+' & = & (25.20 \pm  0.33) \times  10^{-3} \, ,\nonumber \\
\lambda_+'' & = & (12.85 \pm  0.31) \times  10^{-4} \,, \nonumber \\
\lambda_+''' & = & (9.56 \pm  0.28) \times  10^{-5} \, ,
\end{eqnarray}
that compare very well with the recent determinations from $K_{e3}$ decays \cite{Antonelli:2008jg}~:
\begin{eqnarray}
 \lambda_+' & = & (25.20 \pm  0.9) \times  10^{-3} \, ,\nonumber \\
\lambda_+'' & = & (16 \pm  4) \times  10^{-4} \,.
\end{eqnarray}

\end{document}